\documentclass{article}

\usepackage{arxiv}

\usepackage[utf8]{inputenc} 
\usepackage[T1]{fontenc}    
\usepackage{hyperref}       
\usepackage{url}            
\usepackage{booktabs}       
\usepackage{amsfonts}       
\usepackage{nicefrac}       
\usepackage{microtype}      
\usepackage{lipsum}
\usepackage{graphicx}
\graphicspath{ {./images/} }
\usepackage{amsthm}
\theoremstyle{definition}
\newtheorem{definition}{Definition}[section]
\theoremstyle{remark}

\usepackage[english]{babel}
\usepackage{mathtools}
\usepackage{mathabx} 
\usepackage{tabularx}
\title{A HIGHLY  ROBUST  SPARSE  FRACTAL ARRAY }
\author{Kretika Goel \\
  Research Scholar\\
  SENSE\\
  IIT Delhi \\
  \texttt{Kretika.Goel@iddc.iitd.ac.in} \\
   \And
 Monika Aggarwal \\
  Professor\\
  CARE\\
  IIT DELHI \\
  \texttt{maggarwal@care.iitd.ernet.in} \\
  \And
 Subrat Kar \\
  Professor\\
  Dept of Electrical Engg\\
  IIT Delhi \\
  \texttt{subrat@ee.iitd.ac.in}\\
}
\date{}
\begin{document}
\maketitle
\begin{abstract}
The term fractal refers to the fractional dimensions that have recursive nature and exhibit better array factor properties. In this article, we present a new class of sparse array where the recursive nature of a fractal can be used in designing an antenna array called a sparse fractal array by combining sparsity properties of the various sparse array and the recursive nature of fractal array. The most important property of the proposed array is the hole-free difference coarray which makes it a good choice for DOA estimation as various algorithms like coarray MUSIC etc demand hole-free difference coarray. But the performance of any array depends on the presence of essential and nonessential sensors in it which governs whether the difference coarray will get affected upon sensor failure or not. Hence, in this paper, a rigorous analysis is done for various combinations of sparse fractal arrays to test their robustness in presence of a faulty sensor environment.
\end{abstract}


\section{Introduction}
Sparse arrays have attracted considerable attention in various fields such as radar, array signal processing \cite{partan2007survey}, beamforming, the direction of arrival estimation \cite{heidemann2006research}, ultrasound imaging, 5G communications \cite{mao2007wireless}, etc. The important property of sparse arrays i.e. these are the sensor arrays with nonuniform element spacing enables them to resolve more sources than the available physical sensors. This property of sparse arrays arises from the virtual counterpart of the physical array i.e. called a difference coarray which is the pair-wise difference of each sensor location in the physical array. The higher the size of the contiguous region of the difference coarray, the more can be the degree of freedom \cite{yangg2016new}. 

Various types of 1D sparse arrays coined so far include minimum redundancy arrays (MRA) \cite{wang2021fast}, minimum holes arrays (MHA) \cite{qin2019doa}, nested arrays (NA) \cite{pal2010nested} and coprime arrays (CP) \cite{liu2016coprime},\cite{vaidyanathan2010sparse}, CADIS \cite{alamoudi2016sparse}, CACIS \cite{cheng2019doa}, Augmented nested array \cite{ma2020hole}, super nested array \cite{liu2016super}, etc leading to numerous studies proposing diverse sparse configuration. The importance of all these arrays is the increased size of their virtual coarray which makes the DOF of all of them of the order of O(N2) which is high enough as compared to ULA which has DOF O(N) where N is the number of sources present in a physical array. Hence, it can be inferred that the larger the size of the virtual coarray, the more the number of resolvable sources and the better the resolution of the array. Hence, array designing is of prime importance in signal processing because the size of uniform difference coarray is an important metric in achieving better array performance.

In some implementations, arrays with symmetric geometry are highly preferred because they not only ease the computational complexity but also improve the DOA estimation performance \cite{li1993performance}. Moreover such symmetric and recursive geometries also facilitate the array calibration in presence of mutual coupling \cite{ye2009doa}. Hence, In this paper, we consider one such symmetric array as a Fractal array. The introduction of various sparse arrays has sparked great interest in non-uniform arrays, leading to numerous studies. The performance of an antenna array can further be improved by another class of array designing i.e by using fractal geometric techniques. The term fractal, which means irregular fragments, was originally coined by Mandelbrot \cite{werner1999fractal} which has contributed a lot to the new and rapidly growing field of research as fractal antenna engineering which includes the study of fractal-shaped antenna elements, as well as the use of fractals in antenna arrays. 

A wide variety of applications of fractal arrays has been found in fractal electrodynamics \cite{jaggard1991fractal}, in which fractal geometry is combined with electromagnetic theory for the analysis and design of antenna systems. Fractal arrays are recursive arrays that can be formed through the repetitive application of a generating subarray. A generating subarray is a small array at scale one ( P = 1 ) which is used to build larger arrays at higher scales (i.e., P > 1 ). Generating subarray has elements that are turned on and off in a certain pattern. Followed by copying, scaling, and translation to produce the fractal array which can otherwise be considered as a sequence of self-similar subarrays. 

Fractal arrays are of two types: deterministic and natural fractal structures \cite{liu2017maximally} where Deterministic fractal structures are the geometric structures that have exact dimensions for the expansion.  Another type is Random Fractals which are distinguished by the crudest measure of size, which is dimension. All the fractals found in nature are random fractals like mountains, coastal lines, and fractal-shaped flowers. The advantage of using fractal antenna array (FA) is that it improves multi-beam, and multi-band characteristics and also improves array factor behavior. One such example of a Fractal Array is A Cantor array is used for DOA estimation \cite{yang2021extended}. A Cantor array is a recursively generated Array whose difference coarray is hole-free, has a large aperture length, is symmetric, and has a maximal economy. Since resolving more uncorrelated sources using fewer elements is mitigated by Sparse Array by providing a much larger aperture virtual array that resolves more sources.  Hence to combine the properties of sparse arrays and fractal arrays \cite{cohen2020sparse},\cite{raiguru2020doa} a novel sparse fractal array is proposed in this paper which combines these arrays to attain all the properties like extended aperture, high DOF, symmetry, recursiveness, and hole-free virtual coarray, robustness, and array economy.

Sparse arrays are very valuable in terms of cost reduction and effective performance with limited sensors and reduced cost. The array geometry plays important role in the direction of arrival estimation of the sources falling on the array. Hence it should be robust enough to the faulty sensor environment. Faulty sensors change the pattern of difference coarray which leads to failure of coarray MUSIC which is applied to estimate the DOA of the sources falling on the antenna array. 
In general, sensors could fail randomly and may cause the breakdown of the overall system. It is sometimes observed that, for some sparse arrays, such as MRA, sensor failure could shrink the ULA segment in the different coarray which leads to degradation in the performance of the system. According to the literature, the issue of sensor failure can be addressed in two ways \cite{liu2019robustness}. Firstly by developing new algorithms which make the system self-calibrate at times of sensor failure. Secondly by analyzing array geometries in the presence of a faulty sensor environment and then designing the system in such a way that the cost of system designing is governed by the presence of essential and inessential sensors in it \cite{liu2018robustness}. In the literature, the validation of the first method is found in \cite{vigneshwaran2007direction} where DOA estimation is carried out based on a minimal resource allocation network, and in \cite{zhu2015impaired}  array is diagnosed based on Bayesian compressive sensing. However, these papers do not fully exploit the interplay between the array configuration and the exact condition under which these algorithms are applicable. To study the second method various measures are proposed in \cite{alexiou2005investigation} to quantify the robustness of arrays. In \cite{carlin2011robustness} beampattern is studied based on the given sensor failure probability. But the impact of faulty sensors on the difference coarray especially for the sparse arrays still needs to be investigated. This paper tries to do so by examining the proposed sparse fractal arrays in a faulty sensor environment and then compares the results with the already existing most common sparse arrays.

\section{ARRAY  DATA MODEL}
\begin{figure}[ht]
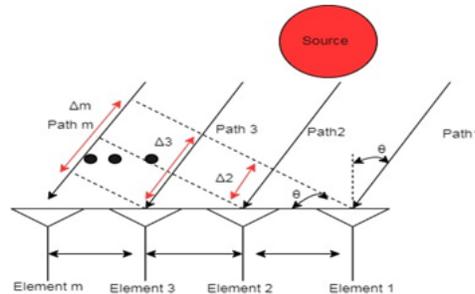

  \begin{center}
  \resizebox{70mm}{!} {\includegraphics *{signalmodel.png}}
     \caption { Array with m Elements}
  \label{fig:signal}
  \end{center}
\end{figure}
Let M narrowband, uncorrelated, and far-field real sources with wavelength $\lambda $ impinge on the 1D antenna array whose elements are located at  n$\lambda $/d  where  n $\in$ L and d is the inter element spacing between the elements as shown in fig.\ref{fig:signal}

The $i^{th}$ source with azimuth angle $\theta $ impinge on the array from the direction \{($\theta_{1} $) ,($\theta_{2} $),($ \theta_{3} $ ) $ \cdots $ ($\theta_{M} $ )\} where $\theta _{i} $  $  \in $  $  [-\pi/2,\pi/2] $

The array output at the time 't' can be expressed as:

\begin{equation}\label{eq:1}
Y(t)= \sum_{i=1} ^{M} A(\theta _{i} ^{'}).S_{i} (t) + N(t) 
\end{equation}

where $\theta _{i} ^{'}$ are normalized DOA such that $\theta _{i} ^{'}$ = (d/ $\lambda$)sin( $ \theta_{i} $) and d=$\lambda /2$ is the inter sensor spacing.
Each element of the steering vector 
$ A(\theta _{i} ^{'} )= [a(\theta _{1}), \cdots, a(\theta _{M}] $ corresponding to the sensors at the location $ n \lambda / d $ where $n \in L $. And is defined as $ e^ {2\pi j \cdot sin(\theta_{i} ^ {'}}$ . Signal $ S _{i}(t) = [s_{1}(t), \cdots, s_{M}(t)]^{T} ]$ is the
source signal vector and N(t) is the white Gaussian noise
vector with zero mean and variance $ \sigma^{2} $. t = 1, 2, • • •, T refers to the sampling time, where T is the total number of snapshots.

The covariance matrix of Y(t) can be written as:
\begin{equation}\label{eq:2}
R = E[Y(t)\cdot Y(t) ^{H}]
    = \sum_{i=1} ^{M}\sigma_{i} ^{2}\cdot A(\theta _{i} ^{'})\cdot A^{H} (\theta _{i} ^{'})  
      + \sigma^{2} I
\end{equation}
where $\sigma_{i} ^{2}$ is the $i^{th}$ source power and $\sigma ^{2} $ is the noise power.

Note that the entity $A(\theta _{i} ^{'})\cdot A^{H} (\theta _{i} ^{'})$ in the covariance matrix defined in eq.\ref{eq:2} is of the form $e^{2\pi(\theta ^{'}) (s_{i}-s_{j})} $ where $(s_{i}-s_{j})$ $\in$ D i.e the difference coarray having $(s_{i}-s_{j})$ as the difference between $i^{th}$ and $j^{th}$ sensor location.

Applying vectorization operation on eq.\ref{eq:2} and reshaping it to get the autocorrelation vector defined on the difference coarray we get,
\begin{equation}
Y_{d}(t) = \sum_{i=1} ^{M}\sigma_{i} ^{2}\cdot A_{D}(\theta _{i} ^{'})  
      + \sigma^{2} I^{'}
\end{equation}

where $ I^{'} =[E^{T}_{1} E^{T}_{2}\cdot E^{T}_{N}]^{T}$ with $E_{i}$ being a column vector of all zeros except value 1 at the $i^{th}$ position. The auto-correlation vector $Y_{d} $, can be considered as sensor output on the difference coarray D. If D contains a contiguous ULA segment $D_{U} $, then DOA can be estimated via coarray MUSIC on the autocorrelation evaluated at $D_{U} $ which ensures that coarray MUSIC can resolve $ O(|L^{2}|) $ uncorrelated sources using $|L|$ physical sensors thereby enhancing the degree of freedom.

Sparse arrays are designed so that the difference-coarray D contains a ULA part around the origin, which is denoted by $D_{U} $. Based on the relationship between D and $D_{U} $, the signal over $D_{U} $ is then constructed as follows:

\begin{equation}
Y_{d_{U}}(t) 
= \sum_{i=1} ^{M}\sigma_{i} ^{2}\cdot A_{D_{U}}(\theta _{i} ^{'})  
      + \sigma^{2} I^{'}
\end{equation}

Doa estimation using finite snapshots can be carried out by calculating the finite snapshot version of Y(t),$Y_{d}(t)$, $Y_{d_{U}}(t)$ and R as $ Y'(k) $ ,$Y'_{d}(k)$, $Y'_{d_{U}}(k)$ and $R'$ respectively where $ k= 1,2,3..... K $ be K realizations of eq.\ref{eq:1}. K is the total number of snapshots. The covariance matrix for which can be estimated as $R' = \sum_{k=1} ^{K} Y'(k).Y'(k) ^{H} /K $ . From R' finite snapshot autocorrelation vector on $D_{U}$ can be formulated as $Y'_{d_{U}}(k)$. Moreover, for sparse array, the covariance matrix R is not hermitian Toeplitz in general but for coarray, we can construct a hermitian Toeplitz matrix by the method explained in \cite{liu2015remarks}.

Now partitioning the signal subspace and noise subspace of this matrix and then applying coarray MUSIC to it will give the desired spectrum of the signal which helps in DOA estimation.

\section{PROPOSED SPARSE FRACTAL ANTENNA ARRAY}
First explaining the basic Cantor arrays [6] which are the fractal arrays that can be defined recursively as :
\begin {equation}
C_{r+1} = C_{r} U ( C_{r} + 3^{r}) where , r \in N
\end{equation}
Note that the cantor array definition is followed from [2] which states that the basic Cantor arrays are symmetric and Cr has N = 2r physical elements. Let C =[0,1] then Cr will have hole-free difference coarray Dr with an aperture as 3r. Its difference coarray is of the order $ O(Nlog_{2}3) \equiv O(N1.585)$ which results in degraded performance when compared to sparse arrays whose difference coarray satisfies O(N2) property. To overcome this limitation of a fractal array we propose a new architecture in which the Cantor array is mixed with the sparse array that will generate an array with increased DOF and reduced RMSE thus showing better performance over other arrays.

\begin{definition}[Proposed 1D SFA]
The proposed SFA(sparse fractal array ) consists of two sub-arrays, i.e. sub-array 1 and sub-array 2. 
Let Sub-array 1 is an M elements sparse array and lets sub-array 2 be an N elements Fractal array, which is created from Cantor set. 
The element position of in subarray 1 and subarray2 is defined by the $\widetilde{S_{1}}$ and $\widetilde{S_{2}}$ where
\begin{equation}
\widetilde{S_{1}} = [m1,m2 ,.....M ]d1,m1 = 1 
\end{equation}
\begin{equation}
    \widetilde{S_{2}} = [n1, n2 ,......N ]d2 , n1 = 0 
\end{equation}
Where m and n are integers and d1 and d2 are inter-element spacing such that d2 = (2M +1)d1.
And the proposed array will be denoted as:

\begin{equation}
    \widetilde{S} = \widetilde{S_{1}} \oplus \widetilde{S_{2}}
\end{equation}

Where element position of proposed SFA is expressed by the cross summation $\oplus$ defined over $\widetilde{S_{1}}$ and $\widetilde{S_{2}}$ and the SFA will have MN elements. 

Example 1:
Let subarray 1 be nested array whose sensor location for M=6 is defined in the set $\widetilde{S_{1}}$. Let d1=1 so the normalized sensor positions in the first subarray are given as :
$\widetilde{S_{1}}$ = [1 2 3 4 8 12].
For constructing $\widetilde{S_{2}}$ we will use the basic Fractal array (FA) ={0,1} in the paper.
Now $\widetilde{S_{2}}$ = [0 1]d2 
where d2 = (2mM +1)d1 =(2*6 +1)d1=13d1
$\widetilde{S_{2}}$ =[0 13]d1
Using the above relation $\widetilde{S_{1DSFA}}$ = $\widetilde{S_{1}}$ $\oplus$ $\widetilde{S_{2}}$ we get,
$\widetilde{S}$=[1 2 3 4 8 12 14 15 16 17 21 25] 
\end{definition}

\begin{definition}[Difference coarray of the proposed SFA]
 The difference coarray $\widetilde{D}$ for an array $\widetilde{S}$ is defined as the set of differences between the sensor locations given in set $\widetilde{S}$. $\widetilde{D}$  = \{n1-n2 : n1,n2  $\in$ $\widetilde{S}$ \}
Let $\widetilde{D_{u}}$ represent the contiguous range of the sensors in the difference coarray $\widetilde{D}$ which is hole free and the most useful when comes to the application of estimation algorithms. The virtual sensor positions in the difference coarray are also called lags and for case1 the virtual aperture of the proposed SFA extends from -24 to 24. In the given example $\widetilde{D_{u}}$  also lies in the range -24 to 24 as there are no holes present in the difference coarray of the proposed SFA. Therefore all information on the difference coarray can be exploited. Fig.\ref{fig:NFA} shows the physical array of the proposed SFA and its different coarray .
 
\end{definition}
\begin{definition}[Weights Function of the proposed SFA]
The weight function w ( k ) can be understood as the number of sensor pairs with separation $k \in Z $ where Z  is an integer. This can also be written as w (k ) = $ | (n 1 , n 2 ) \in L^{2}  : n 1 - n 2 = k |$. By definition, w ( k ) is an  integer-valued even function, i.e. w (-k ) = w (k ) .
Designing any sparse array requires the optimal utilization of the difference between coarray D and the weight function w ( k). For example, both the MRA and the MHA have different coarrays of size $ O(N^2 )$ but the problem in MRA is that it has no close form expression hence the work in \cite{raiguru2019new} and \cite{cohen2019sparse} lacks the closed form expression to deal with sparse and fractal array as the sparse arrays which have closed-form expressions for the sensor locations are simple to compute like nested array, coprime array, generalized coprime array, and super nested array, etc. Hence, we make use of all such sparse arrays whose closed-form expression exists which can be used for designing sparse fractal arrays. Hence, we make use of all such sparse arrays whose closed-form expression exists which can be used for designing sparse fractal arrays. 
\end{definition}

\section{NUMERICAL EXAMPLES}
While designing any array configurations we need to select the basic arrays which fulfill some laid down criteria according to literature first among them is that the sensor locations should be expressed in closed-form. Secondly, the difference coarray of the basic array should be hole free because the estimation performance of various DOA estimation methods, e.g., MUSIC and ESPRIT depends on the cardinality of the central ULA segment of the difference coarray. Hence, for the proper estimation of the number of sources hole free difference coarray is required. Thirdly, the difference coarray should be large enough to achieve increased DOF concerning the number of sensors. So keeping all such criteria in mind various configurations of sparse fractal arrays have been tried and tested using the basic arrays as a nested array, coprime array, Augmented nested arrays, and super nested arrays which are expressed one by one.

\paragraph{1.}[Nested Fractal array ( NFA)]

\begin{figure}[ht]
    \centering
   \resizebox{90mm}{!}{\includegraphics[width=0.48\textwidth]{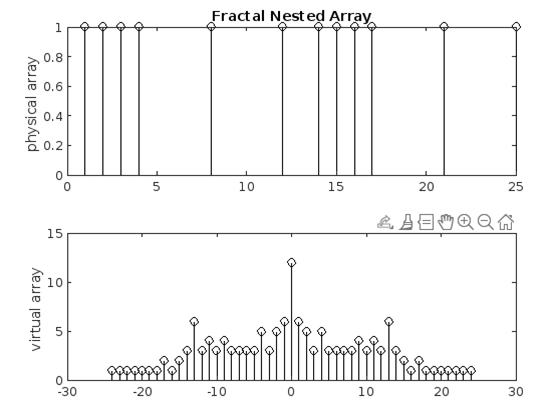}}
   \resizebox{70mm}{!}{\includegraphics[width=0.48\textwidth]{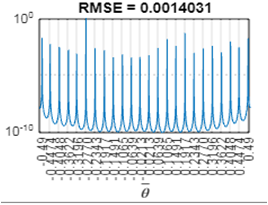}}
   \caption{(a) Physical and Virtual array of NFA,(b) normalized MUSIC Spectrum of 24 signals applied on it for doa estimation along with RMSE.}
    \label{fig:NFA}
\end{figure}

A two level nested array consists of two uniform linear arrays called inner and outer uniform linear arrays where the inner ULA has N1 elements with spacing d1 and the outer ULA has N2 elements with spacing d2 such that d2=(N1 + 1)d1. Sensors locations of the physical array is given by $ \{md1 , (m = 1,2,3 \cdots N1) \cup nd2, (n=1,2,3 \cdots N2)\} $ where if N is even then N1=N2=N/2 and if N is odd then N1=(N-1)/2; N2=(N+1)/2. Using the basic sparse array as a nested array for N=6 and taking the cross sum with the basic fractal array i.e. [0 1] as defined in example 1 we obtain the NFA as shown in Fig along with its difference coarray which has no holes present in it. The central ULA segment of the difference coarray of the proposed NFA is $\widetilde{D_{u}}$ hence it can resolve up to $|(\widetilde{D_{u}}-1 )/2|$ number of sources . Hence, with 6 physical sensor and $|\widetilde{D_{u}}|$ = 49 , it can estimate maximum of 24 sources. Therefore to check the effectiveness let us assume 24 uncorrelated sources falling on the RCPA whose normalized DOA are picked up randomly from $\theta $  = [- 0.5, 0.5] . Let SNR =0 dB and snapshots be 500. Applying MUSIC on the finite snapshot version of the above model, we get the following results as shown in Fig.\ref{fig:NFA} with 0.0014 RMSE achieved. 

\begin{figure}[ht]
    \centering
   \resizebox{90mm}{!}{\includegraphics[width=0.48\textwidth]{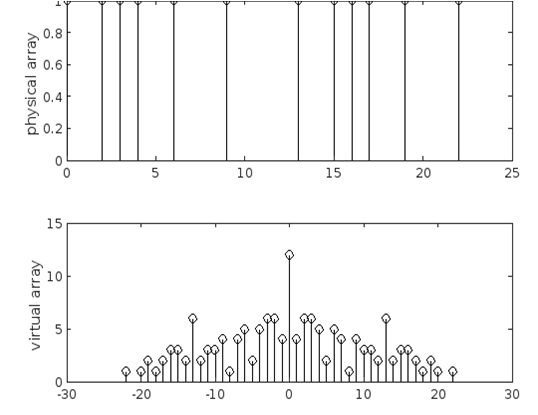}}
   \resizebox{60mm}{!}{\includegraphics[width=0.48\textwidth]{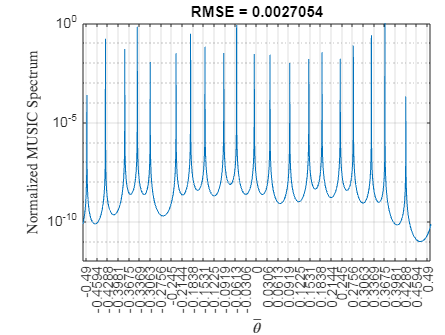}}
    \caption{(a) Physical and Virtual array of CFA,(b) normalized MUSIC Spectrum of 22 signals applied on it for doa estimation along with RMSE. }
    \label{fig:CFA}
\end{figure}

\paragraph{2.}[Coprime Fractal Array(CFA)]
According to the proposed method let subarray 1 be conventional coprime array whose sensor location for coprime pair M=2 and N=3 and total number of elements as $ 2M + N - 1 = 6$ is defined in the set $\widetilde{S_{1}}$ . Let d1=1 so the normalized sensor positions in the first subarray is given as : $\widetilde{S_{1}}$ = [0 2 3 4 6 9].
For constructing $\widetilde{S_{2}}$ we will use the basic Fractal array (FA) = [0,1] in the paper. Now $\widetilde{S_{2}}$ = [0 1]d2, Where d2 = (2mM +1)d1 =(2*6 +1)d1=13d1 $\widetilde{S_{2}}$ 
=[0 13]d1. Using the above relation $\widetilde{S_{1DSFA}}$ = $\widetilde{S_{1}}$ $\oplus$ $\widetilde{S_{2}}$ we get, $\widetilde{S}$=[ 0 2 3 4 6 9 13 15 16 17 19 22] and the virtual aperture of the proposed CFA extends from -24 to 24 . In the given example $\widetilde{D_{u}}$ also lies in the range -22 to 22 also there are no holes present in the difference coarray of the proposed CFA. Let 22 uncorrelated sources fall on the CFA whose normalized DOA are picked up randomly from $\theta $  = [-0.5, 0.5]. Let SNR =0 dB and snapshots be 500. Applying MUSIC on the finite snapshot version of the above model, we get the following results as shown in Fig.\ref{fig:CFA} with 0.0027 RMSE achieved.

\paragraph{3.}[Augmented GenI Fractal Array(AUGGENIFA)]

\begin{figure}[ht]
    \centering
   \resizebox{90mm}{!}{\includegraphics[width=0.48\textwidth]{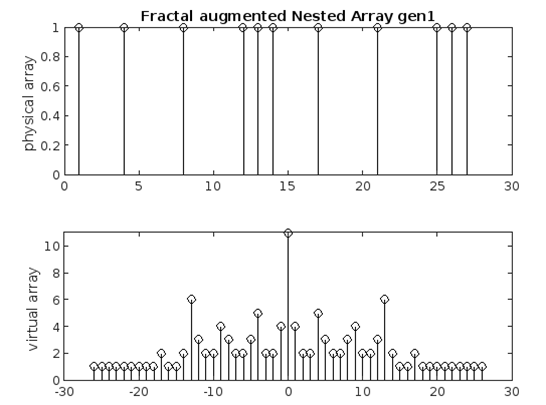}}
   \resizebox{70mm}{!}{\includegraphics[width=0.48\textwidth]{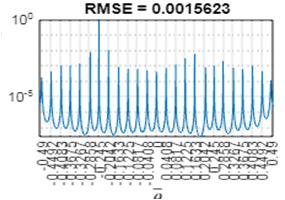}}
    
    \caption{(a) Physical and Virtual array of AUGGENIFA,(b) normalized MUSIC Spectrum of 26 signals applied on it for doa estimation along with RMSE. }
    \label{fig:AUGGENIFA}
\end{figure}

An augmented nested array or ACA was formed by first splitting the dense subarray of the nested array into various parts, which is further rearranged in a particular manner at the sides of the sparse subarray. Using this concept, an augmented nested array (ANA) was designed by splitting the dense ULA of the nested array into a few left/right sub-arrays which are then arranged on either side of the low element density subarray of the nested array. According to the proposed method let subarray 1 be AUGGENI for N=6 a whose sensor location is defined in the set 
$\widetilde{S_{1}}$ . Let d1=1 so the normalized sensor positions in the first subarray is given as : $\widetilde{S_{1}}$ = [1 4 8 12 13 14]. For constructing $\widetilde{S_{2}}$ 
we will use the basic Fractal array (FA) = [0,1] we get, $\widetilde{S}$=[ 1 4 8 12 13 14 17 21 25 26 27], and the difference coarray of the proposed AUGGENIFA extends from -26 to 26 which means that it can estimate a maximum of 26 sources. For checking the array we take the maximum possible sources i.e. 26 uncorrelated sources falling on the AUGGENIFA whose normalized DOA are picked up randomly from $\theta $  = [-0.5, 0.5]. Let SNR =0 dB and snapshots be 500. Applying MUSIC on the finite snapshot version of the above model, we get the following results as shown in Fig.\ref{fig:AUGGENIFA} with 0.00156 RMSE achieved.

\paragraph{4.}[Augmented GenII Fractal Array(AUGGENIIFA)]

\begin{figure}[ht]
    \centering
   \resizebox{90mm}{!}{\includegraphics[width=0.48\textwidth]{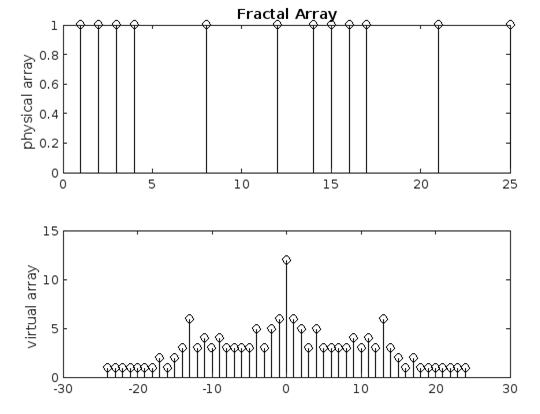}}
   \resizebox{70mm}{!}{\includegraphics[width=0.48\textwidth]{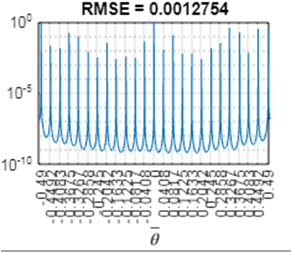}}
    
    \caption{(a) Physical and Virtual array of AUGGENIIFA,(b) normalized MUSIC Spectrum of 26 signals applied on it for doa estimation along with RMSE. }
    \label{fig:AUGGENIIFA}
\end{figure}

Another type of augmented nested array is ANAGENII which is formed by splitting the dense sub-array of the actual nested array into odd and even parts, respectively, and then arranging it on either side of the less dense subarray of the nested array. For N=6, AUGGENII will have the sensor locations at $\widetilde{S_{1}}$ = [1 2 4 8 12 13] and by using basic Fractal array (FA) = [0,1] we get we get, $\widetilde{S}$=[ 1 2 4 8 12 13 14 15 17 21 25 26] . For checking the array we take the maximum possible sources i.e. 26 uncorrelated sources falling on the AUGGENIFA whose normalized DOA are picked up randomly from $\theta $  = [- 0.5, 0.5]. Let SNR =0 dB and snapshots be 500. Applying MUSIC on the finite snapshot version of the above model, we get the following results as shown in Fig.\ref{fig:AUGGENIIFA} with 0.00127 RMSE achieved.

\paragraph{5.}[Super Nested Fractal Array (SNFA)]

\begin{figure}[ht]
    \centering
   \resizebox{90mm}{!}{\includegraphics[width=0.48\textwidth]{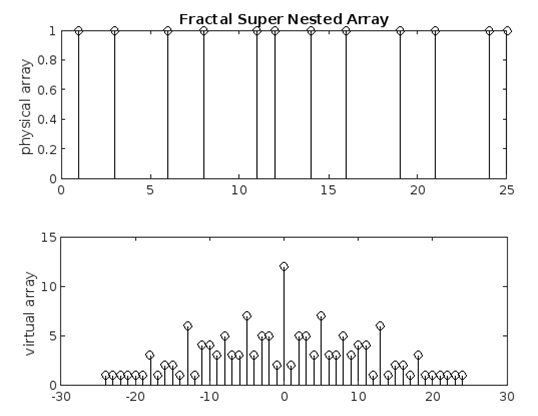}}
   \resizebox{70mm}{!}{\includegraphics[width=0.48\textwidth]{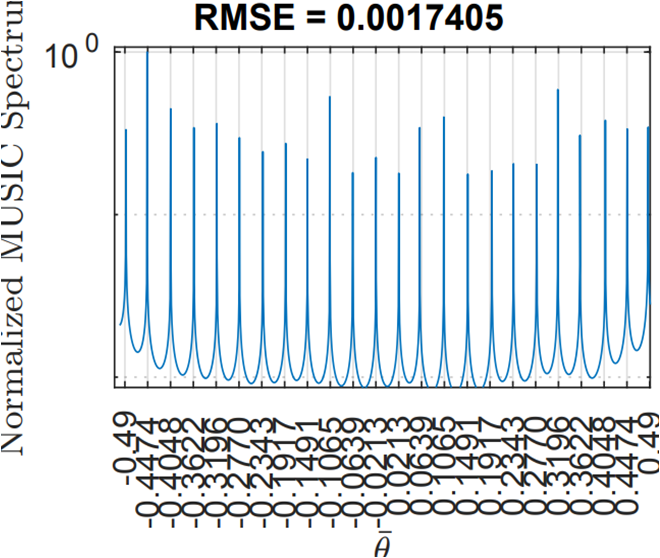}}
    
    \caption{(a) Physical and Virtual array of Supernested Fractal Array,(b) normalized MUSIC Spectrum of 25 signals applied on it for doa estimation along with RMSE. }
    \label{fig:sfa}
\end{figure}

Super nested arrays have the same number of sensors, and the same difference coarray as nested arrays but the arrangement of sensors in a unique manner leads to reduced mutual coupling than nested arrays. For N=6, Super nested arrays will have the sensor locations at $\widetilde{S_{1}}$ = [1 1 3 6 8 11 12] and by using basic Fractal array (FA) = [0,1] we get we get 
SNFA as $\widetilde{S}$=[ 1 3 6 8 11 12 14 16 19 21 24 25] . For checking the validation of the proposed array we take the maximum possible sources i.e. 25 uncorrelated sources falling on the SNFA whose normalized DOA are picked up randomly from $\theta $  = [- 0.5, 0.5]. Let SNR =0 dB and snapshots be 500. Applying MUSIC on the finite snapshot version of the above model, we get the following results as shown in Fig. \ref{fig:sfa} with 0.00174 RMSE achieved.

\section{ROBUSTNESS OF SPARSE FRACTAL ARRAY TO SENSOR FAILURE}

There is a well known theory according to literature that the larger the difference in coarray, the less robust the array is. This can otherwise be stated as given two arrays with the same physical number of elements and the same difference coarray, one could be much more robust as compared to another. Hence, it is necessary to design the sparse array in such a way that it maintains a balance between the size and the robustness of the difference coarray because when such an array is used for doa estimation then the performance is affected both by the size of the different arrays and the robustness properties. 

\begin{definition}[Essential Sensors]
A sensor in an array is said to be essential if its presence and absence makes an effect on the structure of the different coarray. In other words, an essential sensor as its name signifies is essential because its failure can alter the difference coarray which means that it will no longer be hole free. And the presence of holes in the difference coarray means it degrades the performance of estimation algorithms and the array is said to be less robust.
The necessity to study the essentialness property of the sensors is that it helps in system designing because the cost of sensors assignment truly depends on their essential and inessential behavior. For example, let there be two sensors with different costs and qualities. Let the first sensor is costly but has a low failure probability and another sensor is less expensive but easily fails. Therefore, to maintain a balance between the budget and the robustness of an array, the first sensor can be used as an essential sensor, and another sensor can be used as an inessential sensor while designing the array.
\end{definition}

\begin{definition}[ The fragility of a sensor array ]
The term fragility is the characteristic term that is used to define whether the array is robust to sensor failure or not. It can be better stated in terms of fragility i.e. an array is said to be more robust if it is less fragile and vice versa. Hence, the fragility(F) varies from 0 to 1, where 0 means less fragile and more robust and 1 means more fragile and less robust array. The formula for calculating fragility is the number of essential sensors divided by the total number of sensors in an array.
\end{definition}

\begin{figure}[ht]
    \centering
   \resizebox{130mm}{!}{\includegraphics[width=0.48\textwidth]{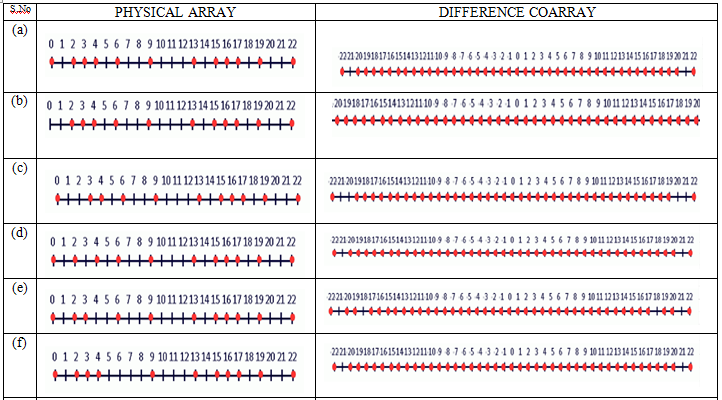}}
   \resizebox{130mm}{!}{\includegraphics[width=0.48\textwidth]{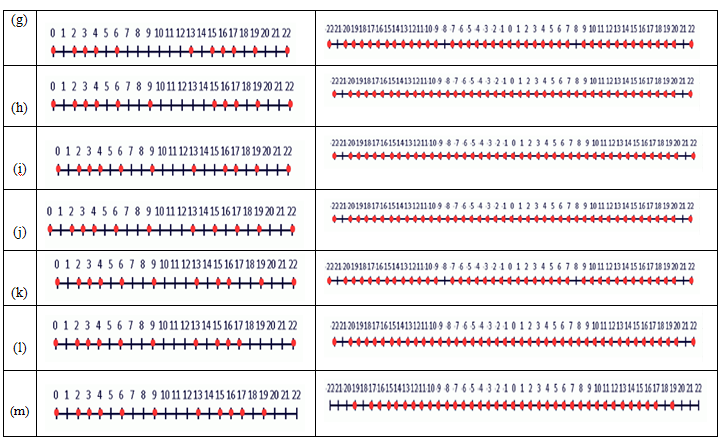}}
   \caption{ (a) The original array and its difference coarray of Proposed Coprime Fractal Array. The array configurations and the difference coarrays after the deletion of (b) the
sensor at 0, (c) the sensor at 2, (d) the sensor at 3, or (e) the sensor at 4,(f) the sensor at 6, (g) the sensor at 9, or (h) the sensor at 13,(i) the sensor at 15, (j) the sensor at 16, or (k) the sensor at 17,(l) the sensor at 19, (m) the sensor at 22 from the original array in (a). Here the sensors are denoted by red dots and empty space resembles holes.}
    \label{fig:RFA}
\end{figure}

\begin{definition}[The R-essentialness property of a sensor array]
The R-essentialness means that it is not always necessary that only one sensor will fail at a time and only its failure will govern the robustness of an array. There can be the possibility that R sensors can fail at a time hence all possibilities need to be considered which define the robustness of an array. All such R possibilities are called R-essentialness. A subarray $\widetilde{Z}$  of $\widetilde{S}$ is said to be k -essential with respect to an array $\widetilde{S}$ if it has the following properties. Firstly, $\widetilde{Z}$   has size exactly R . Secondly, The difference coarray changes when $\widetilde{Z}$  is removed from $\widetilde{S}$ . Note that the R -essentialness is an attribute of a subarray $\widetilde{Z}$ of $\widetilde{S}$ and the fragility for all such subarrays will be called R-Fragility i.e. if two sensors fail at a time then it is 2-essentialness with 2-Fragility defining the robustness of an array.

\end{definition}
Fig.\ref{fig:RFA} explains the concept of essential and inessential sensors i.e the sensors at $\{0,2,4,9,17,22 \}$ are essential sensors because the removal of them alters the difference coarray while the sensors at $\{3,6,13,15,16,19\}$ are inessential sensors as there removal does not alter the original difference coarray of proposed CFA. Hence the total number of essential sensors in CFA are 6 which calculates fragility $F_{1} = 0.5$ which satisfy $0\leq F \leq 1$.Hence, the proposed fractal design is robust to sensor failure.

\section{ SIMULATION RESULTS}
To study the robustness of proposed sparse fractal arrays we consider 1 sensor failure called 1-essentialness, 2 sensor failures called 2-essentialness, and 3 sensor failures at a time as 3-essentialness respectively whose fragility is also calculated as F1, F2, and F3 respectively. The cumulative results are shown in the table but due to space limitation we will only show 1-essentialness results in the table but all F are shown in the Table.\ref{ta} which will be used for further comparisons.

\begin{table}[ht]
\caption{Robustness Analysis of Proposed Sparse Fractal Array}
\begin{center}
\begin{tabular}{|p{1cm}|p{3cm}|p{3cm}|p{3cm}|}
\hline
\textbf{\textit{S.No.}}& \textbf{\textit{Proposed Sparse Fractal Arrays with 12 physical sensors.}}& \textbf{\textit{1-Essentialness }}& \textbf{\textit{Fragility }} \\
\hline
  1	&Nested fractal Array& \{1,2,3,4,17,21,25\}& F1=0.5833 F2=0.8788 F3=0.9909   \\ 
\hline
  2 &Coprime Fractal array&	\{0,2,4,9,17,22\}& F1=0.5000 F2=0.8182 F3=0.9727 \\ 
 \hline
  3&AUGGENI Fractal Array& \{1,4,8,12,17,21,25,26,27\}&	F1=0.8182  F2=1  \\ 
 \hline
  4&AUGGENII Fractal Array&\{1,2 ,3,4,17,21,25\}&	F1=0.5833 F2=0.8788 F3=0.9909 \\ 
\hline
  5&Super Nested Fractal Array&\{1,3,6,8,11,12,21,24,25\}& F1=0.7500 F2=1 \\ 
\hline

\end{tabular}
\label{ta}
\end{center}
\end{table}

After plotting these results of fragility F1 with respect to normalized difference coarray which is obtained by the relation and comparing it with some already existing sparse arrays we get Fig. which shows that out of all proposed fractal arrays coprime fractal array is found to be more robust and very less fragile whereas already existing MRA, MHA, and nested array has F=1 which means they are least robust to sensor failure.
\begin{figure}[ht]
  \begin{center}
  \resizebox{90mm}{!} {\includegraphics *{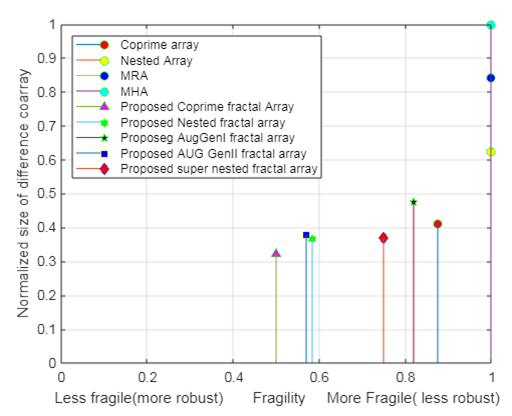}}
     \caption { Robustness analysis for all proposed arrays for F=1 with 12 physical sensors in each sparse fractal array }
  \label{fig:fragility}
  \end{center}
\end{figure}
Fig shows the R fragility of several array configurations with 12 physical sensors and taking 1-essentialness, 2-essentialness, and 3-essentialness with respect to F1, F2, and F3 we found that It can be observed that the ULA is the most robust array since it has the smallest F for all R. The nested array and the super nested array are least robust as $ ( F_{R} = 1 for all 1 \leq R \leq | S | ) $ The prototype coprime array, Proposed NFA, CFA, AUGGENII, AUGGENI are less robust than the ULA but more robust than the nested array and the SNFA.
\begin{figure}[ht]
  \begin{center}
  \resizebox{90mm}{!} {\includegraphics *{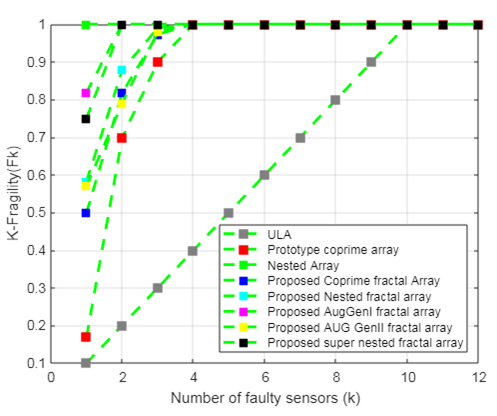}}
     \caption { The R-fragility of several array configurations with 12 physical sensors }
  \label{fig:rfragility}
  \end{center}
\end{figure}
\section{CONCLUDING REMARKS}
In this paper, we have proposed various sparse fractal array configurations and have generated the hole-free difference coarray which is crucial for the functionality of coarray MUSIC and other estimation algorithms. Further, we have studied the robustness of proposed arrays to sensor failure by taking various combinations of sensor failure on the proposed arrays. It helps to study the behavior of the probability that the difference coarray changes when the particular sensor fails. Hence, the essentialness property of the sensors in an array forms the basis for analyzing the robustness of sparse arrays. In the future, it is of considerable interest to design 2D sparse fractal arrays and study the robustness of arrays.


\section*{ACKNOWLEDGMENT}

The authors would like to thank the Center for Sensors, Instrumentation and Cyber-physical Systems Engg (SENSE), Center for Applied Research in Electronics (CARE) and the Department of Electrical Engineering, Indian Institute of Technology Delhi (IIT Delhi) for the facilities used in this research.

\bibliographystyle{unsrt}  
\bibliography{references}  



\end{document}